\documentclass[intlimits,twoside,a4paper]{article}

\usepackage[cp1251]{inputenc}
\usepackage{cmpj3}

\issue{2026}{29}{2}{23503}
\doinumber{10.5488/CMP.29.23503}
\title[Enhanced approach to calculation of cluster integrals for lattice models of matter]%
{Enhanced approach to calculation of cluster integrals for lattice models of matter}
\author[M. V. Ushcats, S. Yu. Ushcats]{
M. V. Ushcats\orcid{0000-0002-0174-1594}\thanks{Corresponding author: \email{mykhailo.ushcats@nuos.edu.ua}}, 
S. Yu. Ushcats\orcid{0000-0002-5250-4505}, 
}
\address{Admiral Makarov National University of Shipbuilding, 9 Prosp. Heroes of Ukraine, Mykolayiv 54025, Ukraine}
\date{Received 19 December 2025; revised 10 February 2026; accepted 10 February 2026; published 29 June 2026}
\begin{document}

\maketitle

\begin{abstract}
The study is devoted to enhancing the existing techniques of calculating Mayer's expansion cluster integrals for lattice models of matter. Two important optimizations are proposed: simplifying the calculation of the integrand at each integration point and reducing the number of such integration points due to eliminating physically identical configurations. Based on those optimizations, new data on high-order cluster integrals are obtained for a number of 2D and 3D lattice models.
\keywords {Mayer's expansion, reducible cluster integral, connected diagram,   lattice statistics, Wheatley's algorithm}
%\printkeywords
%
%\pacs 05.20.-y, 05.50.+q, 05.70.-a, 51.60.+a, 64.60.-i, 75.50.-y
\end{abstract}

\section{Introduction}

Theoretical description of phase transitions in many-particle systems still stays a challenging problem of molecular physics, statistical mechanics, condensed matter physics, physical chemistry, and other related areas. Since the 19th century, the real success has been achieved at the qualitative level only for two extremely specific statistical models: the van der Waals--Maxwell equation of state based on the mean-field approximation of intermolecular interactions \cite{Kac2, Lebowitz2} and Onsager's solution of the Ising problem~\cite{Onsager, Ising}, which permits determining the condensation parameters for the Lee--Yang two-dimensional lattice gas with the square-well potential \cite{LeeYang}. 

Some recent generalizations based on the Mayer cluster expansion \cite{Mayer} have renewed interest to the problem \cite{JML}, but their results are also qualitative rather than quantitative. In particular, an exact generating function for Mayer's expansion in terms of irreducible cluster integrals (virial coefficients) as well as reducible cluster integrals \cite{Mayer} and  new equations of state based on this function \cite{JML, Bannur, PRE5} have finally allowed clearing the long-standing ``problem of the virial expansion divergence'' \cite{Lebowitz1, PRE4, PRE6, UJP4, Pramana} and establishing a strict mathematical saturation-point definition general for classical fluids \cite{JML, PRE6, PRE7, PRE8} and quantum systems~\cite{Suresh2018, Suresh2020}.

In seminal books on statistical mechanics \cite{Huang, Hill, Isihara, Pathria, Gallavotti, Baxter}, the lattice models of matter are considered as thermodynamic systems with a discrete configuration phase-space: lattice gases, the Ising model of magnets, lattice statistics of binary mixtures, etc. Concerning the theory of phase transitions, these statistical models turned out to be of special interest \cite{LeeYang, Gallavotti1968, CMP2, CMP3, PRE3, UJP4, UJP5, PHYSA1, PHYSA2, CMP0} due to a number of reasons. First of all, there exists the above-mentioned expression for phase-transition parameters of the simplest two-dimensional Lee--Yang lattice gas \cite{LeeYang}, which helps with direct comparisons among different analytical or numerical approaches. Secondly, the known ``particle--hole'' symmetry of such models~\cite{PRE3} as well as recent generalization of the lattice statistics \cite{PHYSA1, PHYSA2, CMP0} permit an exact theoretical prediction of the phase-transition activity for any lattice model of gases, magnets, and binary mixtures. The third reason of such interest to the lattice models is that their cluster integrals (i.e., the power coefficients in Mayer's expansion of the partition function \cite{Mayer, JML}) are defined as finite exact sums in contrast to the models with a continuous phase-space, where cluster integrals can only be evaluated by using numerical integration techniques for each value of temperature. Moreover, a special technique of summation \cite{PRE3, PHYSA1} permits defining the lattice statistics cluster integral only once as an exact function of temperature in a form of some polynomial in powers of Mayer's function \cite{Mayer, JML}.

On the other hand, any accurate cluster-based determination of the saturation point (in terms of density and pressure) actually requires information on the corresponding cluster integrals to the orders of hundreds or thousands that, unfortunately, can hardly be determined even by using the modern computational equipment. At the moment, the highest order of the calculated integrals reaches 6--7 for various 2D-lattice models and 5 for 3D models \cite{PRE3, PHYSA1}. Due to a huge number of integration points, any attempt to calculate the integrals of higher orders (and, therefore, to extend the set of exactly known cluster integrals) still turns into a very complex computational problem. 

The present study is devoted to enhancing the algorithm of integration over the lattice configurations in order to calculate the cluster integrals of the orders higher than those of integrals known at the moment. Such enhancing involves two major optimizations: the first one is a fundamental simplification of the Wheatley algorithm \cite{Wheatley} for calculating reducible cluster integrands at each configuration point by moving its most complex stages to the final treatment of the resulting integral and the second is an effective reduction of the actual number of integration points (configurations).

\section{Theoretical background}

\subsection {Simplification of ``per point'' computations}

\begin{figure}[b]
\centering{\includegraphics[scale=0.8]{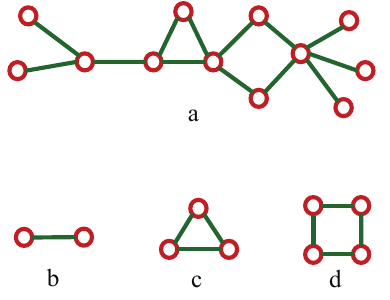}}
\caption{\label{fig1}(Colour online) One of the many connected diagrams representing the $b_{12}$ integrand (a) that, in turn, can be considered as a combination of six 1st-order irreducible integrands (b), one 2nd-order irreducible integrand (c), and one of the parts (d) of the 3rd-order irreducible integrand.}
\end{figure}

In accordance with its definition by the Mayers \cite{Mayer, JML}, the \emph{reducible cluster integral} of the $n$-th order, $b_n$, implies the integration over all the configuration phase-space of $n$ particles. Its integrand is a sum of \emph{all possible} products of pair Mayer's functions, $ f_{ij}  = {f_{ij} \left( {r_{ij}}, T \right)}$, where each particle (say, particle $i$) is ``connected'' by $ f_{ij} $ to another particle ($j$) --- in each product, any particle must be connected at least to one of the rest particles or it may be connected to a number of particles by the corresponding number of Mayer's functions. Graphically, such a single product of Mayer's functions is often represented by the corresponding ``connected diagram'' (or ``connected graph''), where each Mayer's function is called a ``bond'' connecting some pair of particles (see figure~\ref{fig1}a), and the whole reducible integrand is always a sum of all such products possible for the cluster of $n$ particles. 

Obviously, the generation of all possible connected diagrams itself is a very complicated and challenging task for high orders, $n$, even aside from the mentioned many-dimensional integration. As a rule, each reducible cluster integral is considered as a sum of products of relatively simpler \emph{irreducible cluster integrals}, which are directly associated with virial coefficients \cite{Mayer, JML}. There is a great experience in calculating the irreducible integrals (virial coefficients) for various interaction models \cite{KofkePRL, Kofke1, UJP1, UJP2, UJP3, Wheatley, Kofke5, CMP1} at the moment. Except for the simplest first-order irreducible integral (see figure~\ref{fig1}b), the other irreducible integrals (for three, four, and more particles) are represented by ``biconnected diagrams'', where each particle must have at least two ``bonds'' connecting it to the others (see figure~\ref{fig1}c and d). For high-order irreducible integrals, the generation of all the possible diagrams is almost as challenging as for the corresponding reducible integrals. 

In particular, a very effective algorithm to the calculation of the irreducible integrand at a certain integration point was proposed by Wheatley \cite{Wheatley}. Actually, this algorithm may be split into two separate stages: at the first stage, the reducible integrand is determined in a recursive procedure and, at the second stage, the corresponding irreducible integrand is determined in an even more complex recursive procedure. Hereafter, we are interested in calculation of only reducible integrals, because the first stage of Wheatley's algorithm may be essentially simplified, and there are effective procedures for further conversion between reducible and irreducible cluster integrals \cite{PRE4, Pramana, PRE7}. 

The original recursive procedure of computing the reducible integrand, $f_C^{\{ n\} }$, for the $\{n\}$ set of particles at any configuration can be represented by the following relation \cite{Wheatley}:
\begin{equation}
f_C^{\{ n\} } = f_Q^{\{ n\} } - \sum\limits_{\{ s\}  \in \{ n\} } {f_C^{\{ s\} }f_Q^{\{ n\} /\{ s\} }} , \label{eq:1}
\end{equation}
where $f_Q^{\{ m\} }$ is the product of all the possible $\left( f_{ij} + 1 \right)$ for all the pairs in some $\{ m \}$ set of particles. The $\{ s \}$ set always includes the 1st (central) particle (for the set of only this one  central particle, $f_C^1 \equiv 1$). The $\{ n \} / \{ s \}$ set is always complementary to $\{ s \}$  and it never includes the 1st particle (for a set of one particle, $f_Q^{i \ne 1} \equiv 1$). 

That approach requires implementing the recursion of equation~(\ref{eq:1}) (and, hence, computations of all possible $f_C^{\{ s\} }$ and $f_Q^{\{ n\} / \{ s\} }$) at each integration point, although, such ``per point'' computations may be fundamentally simplified. 

A key property of equation~(\ref{eq:1}) is that $f_C^{\{ s\} }$ and $f_Q^{\{ n\} / \{ s\} }$ values of each product in equation~(\ref{eq:1}) are always over independent phase-spaces of separate $\{ s \}$  and $\{ n \} / \{ s \}$  subsets of particles. Therefore, it is possible to integrate the $f_C^{\{ s\} }$ and $f_Q^{\{ n\} / \{ s\} }$ over all the configuration points of the $\{ n \}$ set \emph{separately} --- actually, to take the separate sums for all $\{ s \}$  and $\{ n \} / \{ s \}$ subsets over all the configurations  --- and, only then, to find the resulting sum of the corresponding products after walking over all the configuration points. 
Moreover, as any $\{ s \}$ subset includes a central particle, the correct summation of $f_C^{\{ s\} }$ over the total integration volume must yield the reducible integral of the order $s$ less than $n$, which can actually be already known from previous calculations of reducible integrals of lower orders.

As a result, such an approach eliminates performing a large number of complex recursive computations of $f_C^{\{ s\} }$ values at each configuration point: it requires only collecting the information on $f_Q^{\{ m\} }$ values for certain $\{ m \}$ subsets of $\{ n \}$ and, thus, makes the ``per point'' computations essentially faster.

It should additionally be emphasized that this simplification can be applied to calculating the reducible integrals only. The corresponding irreducible integrals require much more complex ``per point'' computations based on the original Wheatley's algorithm including a huge number of recursions for high~$n$~\cite{Wheatley}. On the other hand, the configuration phase-space of any irreducible integral (and, hence, the total number of relevant integration points) is limited to a greater extent in comparison to the corresponding reducible integral due to the presence of additional ``bonds'' in biconnected diagrams. Therefore, there arises a question, which exactly approach can be more effective,  and  the answer depends on the software and hardware being used for integration. It should be decided whether to perform simple computations over a greater number of points or to perform complex computations over a smaller number of points. Correspondingly, any possibility to reduce the number of integration points would be especially important for reducible integrals.

\subsection {Reducing the integration phase-space}

As it has already been mentioned above, a conventional calculation of $b_n$ implies the integration over the $V^{n-1}$ phase-space volume of $n - 1$ particles around one particle assumed to be the first (or central) particle and $V$ is the maximum volume, where those $n$ particles may form the longest connected diagram (a chain). 

\begin{figure}
\centering{\includegraphics[scale=1]{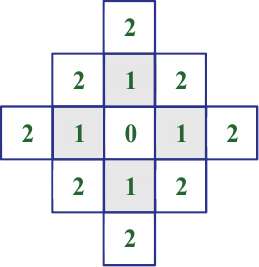}}
\caption{\label{fig2}(Colour online) Integration volume per particle for $n=3$ in the 2D Lee--Yang lattice model. Numbers correspond to coordination spheres.}
\end{figure}

Namely, for the simplest 2D Lee--Yang lattice model \cite{LeeYang}, where each particle can interact with only four its nearest neighbors (in case of $n = 3$, the integration volume, $V$, of this model is shown in figure~\ref{fig2}), the total number of integration points reaches ${\left[ {2n\left( {n - 1} \right) + 1} \right]}^{ n - 1}$. At high orders (see the 1st and 2nd columns of table~\ref{tab:1}), this number makes it almost impossible to perform any real computations over all the integration points and, for other lattice models, the number of relevant configurations becomes even much greater.

\begin{table}[htb]
    \centering{
    \caption{Comparison between non-optimized and enhanced approach for the Lee--Yang model.}
    \label{tab:1}
\vspace{1ex}
\renewcommand{\arraystretch}{1.3}
\noindent{\footnotesize\begin{tabular}{|c|c|c|}
        \hline
        Integral order, & Non-optimized approach, & Enhanced approach,  \\
        $n$ & points & points \\
        \hline
        3 & 169 & 55 \\
        4 & 15 625 & 995 \\
        5 & 2 825 761 & 25 720 \\
        6 & 844 596 301 & 875 946 \\
        7 & 377 149 515 625 & 37 305 195 \\
        8 & 235 260 548 044 817 & 1 916 172 535 \\
        9 & 195 408 755 062 890 625 & 115 608 738 525 \\
        \hline
    \end{tabular}}
\renewcommand{\arraystretch}{1.0}
}
\end{table}

Nevertheless, there is an option to reduce the actual number of integration points. In any reducible integral, all the particles (excluding the central one) are considered to be indistinguishable and, thus, a huge number of integration points may differ only in indices of particles, but they can be physically equivalent in essence.

Therefore, an obvious way to reduce the number of integration points is to make the $n - 1$ particles distinguishable and, then, account for all equivalent configurations (due to permutations of indices) by the corresponding multiplier. In this way, each numbered particle can occupy some limited number of numbered cells belonging to some ``coordination spheres'' (see figure~\ref{fig2}), where it can be connected with ``preceding'' and ``following'' particles of the cluster in their own cells: the first particle always remains in the central cell (sphere~0); the 2nd particle can be placed at the center (sphere~0) and all cells of sphere 1; the 3rd particle can be placed at the same cell as the 2nd one and all the cells after the 2nd (cells after the 2nd in the sphere, where the 2nd stays at the moment and all the following cells of all spheres up to sphere 2); the 4th particle can occupy all the cells beginning from the position of the 3rd one up to ``the last'' cell of sphere 3 and so on.

The integrand of each mentioned configuration must be included to the resulting sum with the integer ``weight'', which is the number of equivalent configurations due to the actual indistinguishability of particles. This weight can easily be determined for each configuration as ${\left( n-1\right) !}/ {\left( {k_2} {k_3} {k_4} {...} {k_n} \right)}$, where ${k_2} = 2$, if the 2nd particle occupies the same cell as the 1st one and ${k_2} = 1$ otherwise; ${k_3} = {k_2} + 1$, if the 3rd particle occupies the same cell as the 2nd one and ${k_3} = 1$ otherwise; ${k_4} = {k_3} + 1$, if the 4th particle occupies the same cell as the 3rd one and ${k_4} = 1$ otherwise, etc.

Let us consider the corresponding algorithm for $n = 3$ (see figure~\ref{fig2}) in detail. Particle~1 is always in the central cell~0 (the sole cell in sphere~0). Particle~2 can occupy the same cell~0 as well as any of 4 cells in sphere~1. When particle~2 occupies the central cell~0, ${k_2} = 2$ and particle~3 can occupy the same cell as well as any of 12 cells in spheres~1 and 2: one position with ${k_3} = {k_2} + 1 = 3$ and 12 positions with ${k_3} = 1$ or 13 positions in total. When particle~2 occupies the first cell of sphere~1, ${k_2} = 1$ and particle~3 can occupy the same first cell of sphere~1 as well as any of 11 other cells in spheres~1 and 2: one position with ${k_3} = {k_2} + 1 = 2$ and 11 positions with ${k_3} = 1$ or 12 positions in total. When particle~2 occupies the second cell of sphere~1, ${k_2} = 1$ and particle~3 can occupy the same second cell of sphere~1 as well as any of 10 ``next'' cells in spheres~1 and 2: one position with ${k_3} = 2$ and 10 positions with ${k_3} = 1$ or 11 positions in total. When particle~2 occupies the third cell of sphere~1, ${k_2} = 1$ and particle~3 can occupy the same third cell of sphere~1 as well as any of 9 next cells in spheres~1 and 2: one position with ${k_3} = 2$ and 9 positions with ${k_3} = 1$ or 10 positions in total. Finally, when particle~2 occupies the fourth cell of sphere~1, ${k_2} = 1$ and particle~3 can occupy the same last cell of sphere~1 as well as any of 8 cells in sphere~2: one position with ${k_3} = 2$ and 8 positions with ${k_3} = 1$ or 9 positions in total. The resulting number of all described configurations is $13 + 12 + 11 + 10 + 9 = 55$ for $n = 3$ (see table~\ref{tab:1}).

For any $n > 2$, the total number of integration points may be expressed as  follows:
\[
N = \sum\limits_{{i_2} = 0}^1 {\sum\limits_{{j_2} = 1}^{s\left( {{i_2}} \right)} {{K_3}\left( {{i_2},{j_2}} \right)} },
\]
where $s \left( i \right)$ is the number of cells in sphere~$i$, which depends on the lattice model, and $K_{k} \left( i, j \right)$ is determined by the recursive relation,
\[
{K_k}\left( {i,j} \right) = \sum\limits_{l = j}^{s\left( i \right)} {{K_{k + 1}}\left( {i,l} \right)}  + \sum\limits_{m = i + 1}^{k - 1} {\sum\limits_{l = 1}^{s\left( m \right)} {{K_{k + 1}}\left( {m,l} \right)} } ,
\]
for $k < n$. Its last value for $k = n$ is
\[
{K_n}\left( {i,j} \right) = s\left( i \right) - j + 1 + \sum\limits_{m = i + 1}^{k - 1} {s\left( m \right)}.
\]
Table~\ref{tab:1} clearly demonstrates the corresponding reduction of the integration points for the above-mentioned 2D Lee--Yang lattice model. For $n = 8$, the proposed enhancement reduces the number of configurations by more than $10^5$ times. For $n = 9$, the reduction is more than $10^6$ times.

Regarding the simplification of Wheatley's algorithm described in the previous subsection, each $f_{Q}^{\{ m \}}$ value must be calculated for only one certain subset $\{ m \}$ of $\{ n \}$ at some integration points in order to walk over the \emph{total phase space of such a subset}. Say, for $b_4$, only the $f_{Q}^{34}$, $f_{Q}^{234}$, $f_{Q}^{1234}$ values must be summarized over all \emph{really distinct} configurations.

\section{Results of computations}

In order to check the effectiveness of both the proposed optimizations, computations of reducible cluster integrals were performed for two 2D square-lattice models (with 4 and 8 neighbor interactions) and two 3D cubic-lattice models (with 6 and 26 neighbor interactions).

In case of nearest-neighbor or next-nearest-neighbor interactions (the square-well interaction potential), Mayer's function, $f_{ij} \left( {r_{ij}}, T \right)$, may take only three certain values: $-1$ (absolute repulsion); some positive $f ( T )$ value (constant finite attraction); $0$ (no interaction). Correspondingly, each $f_{Q}^{\{ m \}}$, as a product of all the $\left( f_{ij} + 1 \right)$ functions, can yield only $0$ (at least, one $ f_{ij} = - 1 $) or ${\left( f + 1 \right)}^k$, where the exponent, $k$, ranges from $0$ to $ m \left( m-1 \right) / 2 $ depending on which $ f_{ij} $ functions are $0$ and which ones are equal to $ f $. This permits to collect only the $k$ values for each non-zero $f_{Q}^{\{ m \}}$ over different configurations and then to calculate the resulting reducible integral, $b_n$, in a form of some exact polynomial in powers of $f ( T )$~\cite{PRE3, PHYSA1}. 

For comparison, the previous non-optimized CPU computations (i.e., computations based mostly on a central processing unit without using the facilities of graphics processing units) were only possible up to the 5th order for the mentioned 2D models and 4th order for 3D ones. In fact, computing the integrals of higher orders required the usage of multithread GPU (graphics processing unit) calculations \cite{Kofke2, PRE3, PHYSA1}. Namely, for $b_6$ and $b_7$ of the Lee--Yang model, the non-optimized multithread GPU computations took minutes and hours, respectively. 

Applying only one of the proposed optimizations --- reducing the number of integration points (but still performing recursive operations based on equation~(\ref{eq:1}) at each point) --- has allowed real computations of the Lee--Yang $b_7$ on CPU: it took about 43 minutes on Intel Core i7-7700. Adding the simplification of Wheatley's algorithm (i.e., avoiding the recursive operations and collecting only the information on the $f_Q^{\{ m\} }$ sets) has additionally reduced the time of such computations of $b_7$ to 4 seconds (about 600 times less)!

Moreover, even CPU computations based on the proposed enhanced approach have provided an important new information on reducible integrals for all the above-mentioned lattice models. In particular, such integrals were calculated up to the 9th order (the $b_9$ computation took about 5 hours) for the Lee--Yang model (2D square model with 4 nearest neighbors, see table~\ref{tab:2}). As to the other models, the achievements are as follows: for 2D square model with 8 neighbors, the reducible integrals were calculated up to the 8th order (see table~\ref{tab:3}); for 3D square model with 6 nearest neighbors --- up to the 7th order (see table~\ref{tab:4}); for 3D square model with 26 neighbors --- up to the 6th order (see table~\ref{tab:5}). 

\begin{table}[h]
	%    \centering
	\caption{Reducible cluster integrals of 2D square lattice model (the Lee--Yang model), where each particle can interact with only four nearest neighbors.}
	\label{tab:2}
	\vspace{1ex}
	\begin{center}
		\renewcommand{\arraystretch}{1.3}
		\noindent{\footnotesize\begin{tabular}{|c|p{0.9\linewidth}|}
				\hline
				%\\[1mm]%
				\rule{0pt}{3mm}
				$n$ & $n {b_n} {\rho_0^{n-1}}$ \\
				%\\[1mm]%
				\hline
				%        1 & 1 \\
				%\\[1mm]%
				2 & $ -1 + 4 f$ \\
				%\\[1mm]%
				\hline
				3 & $ 1 - 12 f + 18 f^{2}$ \\
				%\\[1mm]%
				\hline
				4 & $-1 + 24 f - 100 f^{2} + 88 f^{3} + 4 f^{4}$ \\
				%\\[1mm]%
				\hline
				5 & $ 1 - 40 f + 320 f^{2} - 740 f^{3} + 415 f^{4} + 40 f^{5}$ \\
				%\\[1mm]%
				\hline
				6 & $ -1 + 60 f - 780 f^{2} + 3352 f^{3} - 4986 f^{4} + 1872 f^{5} + 324 f^{6} + 12 f^{7}$ \\
				%\\[1mm]%
				\hline
				7 & $ 1 - 84 f + 1610 f^{2} - 11004 f^{3} + 30093 f^{4} - 31444 f^{5} + 7574 f^{6} + 2184 f^{7} + 154 f^{8}$ \\
				%\\[1mm]%
				\hline
				8 & $ -1 + 112 f - 2968 f^{2} + 29456 f^{3} - 126740 f^{4} + 243472 f^{5} - 186104 f^{6} + 24912 f^{7} + 13364 f^{8} + 1552 f^{9} + 48 f^{10}$ \\
				\hline
				%\\[1mm]%
				9 & $ 1 - 144 f + 5040 f^{2} - 68376 f^{3} + 423342 f^{4} - 1272348 f^{5} + 1816362 f^{6} - 1034496 f^{7} + 36981 f^{8} + 73440 f^{9}$ \\ $ $& $ + 12582 f^{10} + 756 f^{11} + 9 f^{12}$ \\
				\hline
		\end{tabular}}
		\renewcommand{\arraystretch}{1.0}
	\end{center}
\end{table}

\begin{table}[h]
    \centering
    \caption{Reducible cluster integrals of 2D square lattice model, where each particle can interact with all eight of its nearby neighbors.}
    \label{tab:3}
\vspace{1ex}
\renewcommand{\arraystretch}{1.3}
    \noindent{\footnotesize\begin{tabular}{|c|p{0.9\linewidth}|}
        \hline
        $n$ & $n {b_n} {\rho_0^{n-1}}$ \\
        \hline
%        1 & 1 \\
        2 & $ -1 + 8 f$ \\
        \hline
        3 & $ 1 - 24 f + 84 f^{2} + 12 f^{3}$ \\
        \hline
        4 & $ -1 + 48 f - 456 f^{2} + 912 f^{3} + 336 f^{4} + 56 f^{5} + 4 f^{6}$ \\
        \hline
        5 & $ 1 - 80 f + 1440 f^{2} - 7700 f^{3} + 9250 f^{4} + 6360 f^{5} + 2140 f^{6} + 440 f^{7} + 45 f^{8}$ \\
        \hline
        6 & $ -1 + 120 f - 3480 f^{2} + 34448 f^{3} - 118452 f^{4} + 78912 f^{5} + 98460 f^{6} + 52020 f^{7} + 18006 f^{8} + 4140 f^{9}+ 564 f^{10} $ \\ $ $& $ + 36 f^{11}$ \\
        \hline
        7 & $ 1 - 168 f + 7140 f^{2} - 111776 f^{3} + 712278 f^{4} - 1667288 f^{5} + 400036 f^{6} + 1292984 f^{7} + 995533 f^{8} + 491680 f^{9}$ \\ $ $& $ + 173446 f^{10} + 43624 f^{11} + 7546 f^{12} + 812 f^{13} + 42 f^{14}$ \\
        \hline
        8 & $ -1 + 224 f - 13104 f^{2} + 296352 f^{3} - 2965896 f^{4} + 13198336 f^{5} - 21261568 f^{6} - 3432576 f^{7} + 14030832 f^{8}$ \\ $ $& $ + 16053536 f^{9} + 10632400 f^{10} + 5076192 f^{11} + 1830276 f^{12}+ 500464 f^{13} + 101152 f^{14} + 14240 f^{15}  $ \\ $ $& $+ 1232 f^{16} + 48 f^{17}$ \\
        \hline
    \end{tabular}}
\renewcommand{\arraystretch}{1.0}
\end{table}

Thus, the number of exactly known reducible integrals has been increased by 1--2 orders for the considered lattice models. By using the recursive procedure of conversion \cite{PRE4, Pramana, PRE7}, the corresponding irreducible cluster integrals have also been calculated in the same form of exact polynomials in powers of Mayer's function $f (T)$ (see tables~\ref{tab:6}--\ref{tab:9}).

\begin{table}[h]
	\centering
	\caption{Reducible cluster integrals of 3D cubic lattice model, where each particle can interact with only six nearest neighbors.}
	\label{tab:4}
	\vspace{1ex}
	\renewcommand{\arraystretch}{1.3}
	\noindent{\footnotesize\begin{tabular}{|c|p{0.85\linewidth}|}
			\hline
			$n$ & $n {b_n} {\rho_0^{n-1}}$ \\
			\hline
			%        1 & 1 \\
			2 & $ -1 + 6 f$ \\
			\hline
			3 & $ 1 - 18 f + 45 f^{2}$ \\
			\hline
			4 & $ -1 + 36 f - 246 f^{2} + 380 f^{3} + 12 f^{4}$ \\
			\hline
			5 & $ 1 - 60 f + 780 f^{2} - 3100 f^{3} + 3330 f^{4} + 240 f^{5}$ \\
			\hline
			6 & $ -1 + 90 f - 1890 f^{2} + 13776 f^{3} - 37197 f^{4} + 29574 f^{5} + 3732 f^{6} + 108 f^{7}$ \\
			\hline
			7 & $ 1 - 126 f + 3885 f^{2} - 44632 f^{3} + 216300 f^{4} - 432012 f^{5} + 259784 f^{6} + 50400 f^{7} + 3150 f^{8} + 56 f^{9}$ \\
			\hline
	\end{tabular}}
	\renewcommand{\arraystretch}{1.0}
\end{table}

\begin{table}[h]
	\centering
	\caption{Reducible cluster integrals of 3D cubic lattice model, where each particle can interact with all twenty six of its nearby neighbors.}
	\label{tab:5}
	\vspace{1ex}
	\renewcommand{\arraystretch}{1.3}
	\noindent{\footnotesize\begin{tabular}{|c|p{0.85\linewidth}|}
			\hline
			$n$ & $n {b_n} {\rho_0^{n-1}}$ \\
			\hline
			%        1 & 1 \\
			2 & $ -1 + 26 f$ \\
			\hline
			3 & $ 1 - 78 f + 975 f^{2} + 132 f^{3}$ \\
			\hline
			4 & $ -1 + 156 f - 5226 f^{2} + 41844 f^{3} + 14832 f^{4} + 2760 f^{5} + 268 f^{6}$ \\
			\hline
			5 & $ 1 - 260 f + 16380 f^{2} - 333380 f^{3} + 1899895 f^{4} + 1210380 f^{5} + 470360 f^{6} + 134660 f^{7} + 27495 f^{8}$ \\ $ $& $ + 3760 f^{9} + 280 f^{10}$ \\
			\hline
			6 & $ -1 + 390 f - 39390 f^{2} + 1453592 f^{3} - 20680311 f^{4} + 87818118 f^{5} + 86340444 f^{6} + 52241940 f^{7} + 24851418 f^{8}$ \\ $ $& $ + 9659124 f^{9} + 3078828 f^{10} + 795192 f^{11} + 162486 f^{12} + 25416 f^{13} + 2808 f^{14} + 168 f^{15}$ \\
			\hline
	\end{tabular}}
	\renewcommand{\arraystretch}{1.0}
\end{table}

\begin{table}[h]
    \centering
    \caption{Irreducible cluster integrals of 2D square lattice model (the Lee--Yang model), where each particle can interact with only four nearest neighbors.}
    \label{tab:6}
\vspace{1ex}
\renewcommand{\arraystretch}{1.3}
    \noindent{\footnotesize\begin{tabular}{|c|p{0.85\linewidth}|}
        \hline
        $k$ & $k {\beta _k} {\rho_0^k}$ \\
        \hline
        1 & $ -1 + 4 f$ \\
        \hline
        2 & $ -1 - 12 f^{2}$ \\
        \hline
        3 & $ -1 + 12 f^{2} + 40 f^{3} + 12 f^{4}$ \\
        \hline
        4 & $ -1 - 80 f^{3} - 220 f^{4} - 160 f^{5}$ \\
        \hline
        5 & $ -1 + 40 f^{3} + 600 f^{4} + 1704 f^{5} + 1380 f^{6} + 60 f^{7}$ \\
        \hline
        6 & $ -1 - 588 f^{4} - 5376 f^{5} - 13440 f^{6} - 10584 f^{7} - 1428 f^{8}$ \\
        \hline
        7 & $ -1 + 196 f^{4} + 7504 f^{5} + 47880 f^{6} + 102624 f^{7} + 82236 f^{8} + 19824 f^{9} + 336 f^{10}$ \\
        \hline
        8 & $ -1 - 4896 f^{5} - 84720 f^{6} - 406944 f^{7} - 785916 f^{8} - 663552 f^{9} - 216432 f^{10} - 11232 f^{11} + 72 f^{12}$ \\
        \hline
    \end{tabular}}
\renewcommand{\arraystretch}{1.0}
\end{table}

\begin{table}[h]
    \centering
    \caption{Irreducible cluster integrals of 2D square lattice model, where each particle can interact with all eight of its nearby neighbors.}
    \label{tab:7}
\vspace{1ex}
\renewcommand{\arraystretch}{1.3}
    \noindent{\footnotesize\begin{tabular}{|c|p{0.85\linewidth}|}
        \hline
        $k$ & $k {\beta _k} {\rho_0^k}$ \\
        \hline
        1 & $ -1 + 8 f$ \\
        \hline
        2 & $ -1 - 24 f^{2} + 24 f^{3}$ \\
        \hline
        3 & $ -1 + 24 f^{2} - 64 f^{3} - 144 f^{4} + 168 f^{5} + 12 f^{6}$ \\
        \hline
        4 & $ -1 + 80 f^{3} + 680 f^{4} - 1280 f^{5} - 1760 f^{6} + 1120 f^{7} + 180 f^{8}$ \\
        \hline
        5 & $ -1 - 40 f^{3} - 1320 f^{4} + 4728 f^{5} + 17520 f^{6} - 8940 f^{7} - 17580 f^{8} + 8460 f^{9} + 2820 f^{10} + 180 f^{11}$ \\
        \hline
        6 & $ -1 + 1176 f^{4} - 9744 f^{5} - 78960 f^{6} + 17976 f^{7} + 218736 f^{8} - 87192 f^{9} - 216636 f^{10}$ \\ $ $& $ + 41664 f^{11} + 32844 f^{12} + 4872 f^{13} + 252 f^{14}$ \\
        \hline
        7 & $ -1 - 392 f^{4} + 11312 f^{5} + 198296 f^{6} + 61328 f^{7} - 1273048 f^{8} + 337680 f^{9} + 3534384 f^{10}$ \\ $ $& $ + 101136 f^{11} - 2229108 f^{12} - 7056 f^{13} + 312368 f^{14} + 80864 f{^15} + 8624 f^{16} + 336 f^{17}$ \\
        \hline
    \end{tabular}}
\renewcommand{\arraystretch}{1.0}
\end{table}

\begin{table}[h!]
    \centering
    \caption{Irreducible cluster integrals of 3D cubic lattice model, where each particle can interact with only six nearest neighbors.}
    \label{tab:8}
\vspace{1ex}
\renewcommand{\arraystretch}{1.3}
    \noindent{\footnotesize\begin{tabular}{|c|p{0.85\linewidth}|}
        \hline
        $k$ & $k {\beta _k} {\rho_0^k}$ \\
        \hline
        1 & $ -1 + 6 f$ \\
        \hline
        2 & $ -1 - 18 f^{2}$ \\
        \hline
        3 & $ -1 + 18 f^{2} + 60 f^{3} + 36 f^{4}$ \\
        \hline
        4 & $ -1 - 120 f^{3} - 450 f^{4} - 480 f^{5}$ \\
        \hline
        5 & $ -1 + 60 f^{3} + 1170 f^{4} + 4356 f^{5} + 4620 f^{6} + 540 f^{7}$ \\
        \hline
        6 & $ -1 - 1134 f^{4} - 13104 f^{5} - 41580 f^{6} - 44856 f^{7} - 11340 f^{8} + 336 f^{9}$ \\
        \hline
    \end{tabular}}
\renewcommand{\arraystretch}{1.0}
\end{table}

\begin{table}[h!]
    \centering
    \caption{Irreducible cluster integrals of 3D cubic lattice model, where each particle can interact with all twenty six of its nearby neighbors.}
    \label{tab:9}
\vspace{1ex}
\renewcommand{\arraystretch}{1.3}
    \noindent{\footnotesize\begin{tabular}{|c|p{0.85\linewidth}|}
        \hline
        $k$ & $k {\beta _k} {\rho_0^k}$ \\
        \hline
        1 & $ -1 + 26 f$ \\
        \hline
        2 & $ -1 - 78 f^{2} + 264 f^{3}$ \\
        \hline
        3 & $ -1 + 78 f^{2} - 1324 f^{3} + 3312 f^{4} + 8280 f^{5} + 804 f^{6}$ \\
        \hline
        4 & $ -1 + 2120 f^{3} - 22990 f^{4} - 36000 f^{5} + 277360 f^{6} + 399280 f^{7} + 109980 f^{8} + 15040 f^{9} + 1120 f^{10}$ \\
        \hline
        5 & $ -1 - 1060 f^{3} + 52410 f^{4} - 16524 f^{5} - 2373840 f^{6} + 1540140 f^{7} + 20301180 f^{8} + 25901040 f^{9}$ \\ $ $& $ + 12469740 f^{10} + 3757560 f^{11} + 812430 f^{12} + 127080 f^{13} + 14040 f^{14} + 840 f^{15} $ \\
        \hline
    \end{tabular}}
\renewcommand{\arraystretch}{1.0}
\end{table}

\begin{figure}
\centering{
\includegraphics[scale=0.9]{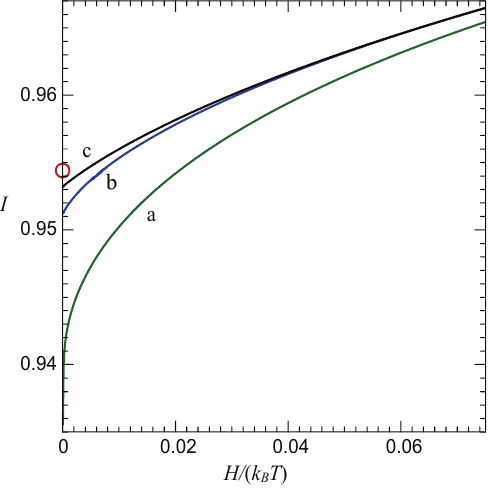}
}
\caption{\label{fig3}(Colour online) Magnetization curves of the 2D square Ising model with nearest-neighbor interactions at $T = 0.8 T_{\rm C}$ ($T_{\rm C}$ is the Curie point). The circle corresponds to the exact spontaneous magnetization point of this Lee--Yang model \cite{Onsager, LeeYang} at the same $T$. For each curve, Mayer's expansion includes 10000 reducible cluster integrals calculated by using the scaling technique \cite{UJP5, PHYSA2, CMP0} on the basis of some limited set of irreducible integrals: a --- $\{ \beta_1, \beta_2 \}$; b --- $\{ \beta_1, \beta_2, \beta_3, \beta_4, \beta_5, \beta_6 \}$; c --- $\{ \beta_1, \beta_2, \beta_3, \beta_4, \beta_5, \beta_6, \beta_7, \beta_8 \}$.}
\end{figure}

In turn, the obtained new information permits to significantly improve the theoretical description of subcritical behavior (including the phase transitions) for lattice models of gases \cite{PRE4, PRE7, PRE8}, the Ising models of magnets \cite{PHYSA1}, and lattice models of binary mixtures \cite{PHYSA2, CMP0}. For example, figure~\ref{fig3} demonstrates an obvious improvement of theoretical magnetization curves, $ I ( H )$, calculated for the Lee--Yang magnetic on the basis of Mayer's expansion for the Ising model \cite{PHYSA1}. In that Mayer's expansion, the set of 10000 power coefficients (i.e., reducible integrals) is approximated \cite{UJP5, CMP0} by using the information on some limited number of exactly known irreducible cluster integrals.

Such scaling technique, which permits to approximate almost unlimited $\{ b_n \}$ sets on the basis of some finite $\{ \beta _k \}$ set, is well described in a number of papers \cite{UJP5, PHYSA2, CMP0}. In brief, the actual value of phase-transition activity, $z_S$, which is always exactly known for any lattice model, strongly determines the asymptotic behavior of high-order reducible integrals \cite{Pramana}. On the other hand, a certain $\{ \beta _k \}$ set, which is always limited at some $k_{\rm max}$ order, determines the correspondingly incorrect value of phase-transition activity, $z^{\rm incorrect}_{S}$, as well as any unlimited $\{ b_n \}$, where only low-order reducible integrals are correct up to $n_{\rm max} = k_{\rm max} + 1$ and all the others are incorrect. In order to ``correct'' such incorrect integrals of the orders higher than $n_{\rm max}$ and reproduce their correct asymptotics at very high orders, they can simply be scaled by the ratio of correct and incorrect values of phase-transition activity in appropriate power:
\[
b_{n} = {b^{\rm incorrect}_{n}} {\left( {\frac{z^{\rm incorrect}_S}{z_{S}}} \right)}^{n-1} .
\]

Figure~\ref{fig3} clearly demonstrates that, as the order, $k_{\rm max}$, of the considered irreducible integrals is increased, the corresponding magnetization curves at $H \to 0$ come closer to the spontaneous magnetization point exactly known for this 2D model \cite{Onsager, LeeYang} (note that the scales on both axes of $I$ and $H$ are considerably enlarged in the figure). Such clearly observed improvement justifies the efforts taken for enhancing the computations of cluster integrals.

It can also be added that applying another and more complex scaling techniques \cite{PRE7} almost does not affect the presented magnetization curves at the scale of figure~\ref{fig3} at least.

\section{Conclusions}

Certain peculiarities of the lattice models of matter (thermodynamic systems with a discrete configuration phase-space) provide two important capabilities to essentially optimize the computations of their cluster integrals: i) a simplification of Wheatley's algorithm \cite{Wheatley}  fundamentally reduces the number of mathematical operations at each integration point (configuration point) by moving the most complex stages of this algorithm to the final treatment of the resulting integral; ii) a particular approach to searching the really different configurations effectively reduces the total number of integration points (see table~\ref{tab:1}).

As a result, the overall enhancement involving both the mentioned optimizations has permitted to extend the information on cluster integrals by one-two orders for a number of different lattice models (see tables~\ref{tab:2}--\ref{tab:5} for reducible cluster integrals and tables~\ref{tab:6}--\ref{tab:9} for irreducible cluster integrals) and, thus, to considerably improve the theoretical description of subcritical behavior (including the phase transitions) for the corresponding lattice gases, magnets (see figure~\ref{fig3} for example), and binary mixtures.

In prospect, there are possible directions of further developing and applying the proposed approach. At the moment, only CPU computations have been performed by using this approach though the essential simplification of per-point operations makes it especially effective for multithread GPU computations. Correspondingly, it seems quite expectable to obtain the information on cluster integrals of higher orders by implementing the same enhanced approach in future GPU computations. Moreover, there is a possibility to apply it to studying one of still challenging problems of Mayer's cluster expansion --- the real dependence of cluster integrals on the integration volume \cite{PRE5, Kofke2023} --- especially in view of a recently proposed approach~\cite{Kofke2023} to their integration by using periodic boundary conditions.

\section*{Acknowledgements}

This work is supported by Science \& Technology Center in Ukraine, Grant NSEP F005 1026. The authors are also grateful to David Kofke and Andrew Schultz for cooperation and express additional thanks to the Referees for their very professional reviews and useful critique.

\newpage

%% Type in your references using {thebibliography} environment 
%% or create them from your bibtex database using cmpj.bst style.

\bibliographystyle{cmpj}
\bibliography{mainRev3}

\newpage

%
%% If you have problems with typesetting in ukrainian uncomment lines below.
%
%  \lastpage
%  \end{document}

\ukrainianpart

\title{Вдосконалений підхід до розрахунків групових інтегралів для ґраткових моделей речовини}
\author{М. В. Ушкац, С. Ю. Ушкац}
\address{Національний університет кораблебудування ім. адмірала Макарова, Просп. Героїв України, 9, Миколаїв, 54025, Україна}
%
%% якщо автор є один або автори є з однієї установи:
%
%  \author{1й Автор, 2й Автор, \ldots}
%  \address{Інститут\ldots}
%
%%

\makeukrtitle

\begin{abstract}
\tolerance=3000%
Дослідження присвячено вдосконаленню існуючих методів обчислення групових інтегралів в розкладі Майєра для ґраткових моделей речовини. Запропоновано два можливих шляхи суттєвої оптимізації таких обчислень: спрощення обчислень в кожній окремій точці та зменшення кількості точок інтегрування завдяки виключенню фізично еквівалентних конфігурацій. На основі цієї оптимізації отримано нову інформацію про групові інтеграли високих порядків для декількох 2D і 3D ґраткових моделей.
\keywords {розклад Майєра, звідні групові інтеграли, зв'язані діаграми, статистика ґраток, алгоритм Уітлі}

\end{abstract}

\end{document}